# Comparative Molecular Dynamics Study of Initial Oxide Film Growth on Al, Mg, and Ti Using a Universal Neural Network Potential

**Yoyo Hinuma** (日沼 洋陽)[1*] **and Taisuke Hasegawa** (長谷川 太祐)[2]

[1] Department of Energy and Environment, National Institute of Advanced Industrial Science and Technology (AIST), 1-8-31, Midorigaoka, Ikeda, Osaka 563-8577, Japan
[2] Matlantis Corporation, Otemachi Building, 1-6-1 Otemachi, Chiyoda-ku, Tokyo, 100-0004, Japan

*y.hinuma@aist.go.jp

## Abstract

Spontaneous formation of an oxide film, which may be passive, is found in many metals such as Al and Ti. We report direct computational observation of oxide film formation on the three major light structural metals, Al, Mg, and Ti, with molecular dynamics (MD) using a universal neural network potential (NNP). $O_2$ molecules were supplied during MD simulations to replenish O consumed by the metal. Al and Ti formed a 1-2 nm thick film within a few ps of the simulation with $O_2$ gas pressure of $10^2$-$10^3$ atm. The film thicknesses of all metal oxides continued to increase proportionally to the logarithm of simulation time, consistent with Cabrera-Mott theory. Al and Ti oxides grow at the surface, while Mg oxide grows into the Mg metal. Al, Mg, and Ti are approximately 4.4, 4.3, and 5.2-fold coordinated, respectively, in the oxide.

## 1. Introduction

Many metals form a surface oxide film immediately upon contact with $O_2$ gas. In particular, the surface oxidation state plays a key role in the three major light structural metals, which are Al, Mg, and Ti. Spontaneous formation of a passive film, or passivation, resists corrosion [1, 2], where corrosion is "deterioration of a material and its desired properties by the environment" [3]. Surface oxidation is a complex process including oxygen adsorption at the surface, growth of oxide islands, and formation of a macroscopically thick oxide layer.[4]

Spontaneous formation of the passive oxide film on Al has been studied extensively. Although the film thickness changes with temperature and O pressure [5], the reported thickness is consistently a few nm. The limiting film thickness of Al, at room temperature, ranges from 2.32 Å at $10^{-8}$ Torr ($1.3\times10^{-6}$ Pa) to saturation at 12.42 Å at 1 Torr ($1.3\times10^{2}$ Pa) based on X-ray photoelectron spectroscopy (XPS) measurements. [6] Another XPS study showed that a film with thickness of ~3.5 nm and O/Al ratio of roughly 1.65 forms over an oxidation time of 300 minutes. [7] The oxide film thickness on Al(431) ranges from ~0.5 nm at 373 K to ~6 nm at 773 K at O pressure of $1.33\times10^{-4}$ Pa after ~4 hours [8]. Hollow particles formed when Al nanoparticles with size less than 6-8 nm were exposed to O, and ~1.5 nm-thick films formed on nanoparticles with a mean diameter of ~20 nm at 423 K, according to transmission electron microscopy (TEM). [9] Medium energy ion scattering showed, for Al(110), 1.5 nm thick oxide formation at 573 K and formation of a rough surface at least 2.2 nm thick at 673 K. [10]

A comparable or thicker passive oxide film spontaneously forms on Ti. A neutron reflectometry study finds that a 2.9±0.3 nm oxide film formed, in air, on a 20 nm-thick Ti film at 293 K. The film thickness increased to 9.4±0.5 nm at 573 K, and using a thinner 5 nm film also increased the thickness to 6.1±0.3 nm at 293 K. [11] The spontaneous oxide film is practically too thin, so anodizing is often used in Ti alloys, which uses a large electric field to drive formation of a thick passivation layer. [12]

The native oxide film thickness on pure Mg in air is reported as 2.6 nm, though the film is not protective and allows corrosion of ~50 mg/cm$^2$/day. [13] Cathodic impurities, such as Fe, Cu, and Ni, speed up anodic dissolution of Mg and thereby enhances corrosion. Alloying that prevents electrochemical degradation is the leading strategy to improve the corrosion resistance of Mg,[14] and coating is an additional important measure to suppress corrosion of Mg.[15]

Cabrera-Mott theory on initial surface oxide film growth states that electrons at the Fermi energy of the metal move toward the oxygen adsorbed at the surface, forming negatively charged oxygen species. As long as the electron level of O is lower than the metal Fermi energy, a potential difference (Mott potential) facilitates movement of electrons toward the surface, resulting in very fast oxidization. The potential disappears above a certain oxide thickness, which slows down the oxidation rate. [16-18] Cabrera and Mott discussed Al and Cu as systems to which their theory is applicable. [16]

Two types of film growth behavior are based on Cabrera-Mott theory. The inverse logarithmic growth law assumes that ion transport is controlled by the electric field across the film. The current density $i$ is related to the electric field strength $E$ as

$i = A\exp(BF)$ (1)

using constants $A$ and $B$, time $t$, and

$F = \Delta U/X$ (2)

where $\Delta U$ is the electrostatic potential difference between two oxide interfaces and $X$ is the film thickness. This $\Delta U$ is considered a constant. The time dependence of the film thickness is obtained as

$1/X = C - D\log(t)$, (3)

where $C$ and $D$ are constants and $t$ is time. [1]

On the other hand, the direct logarithmic growth law suggests that the oxide formation is governed by an activation energy $W$ that linearly increases with $X$. Solving the differential equation

$\mathrm{d}X/\mathrm{d}t = \exp(-W_0 + \mu X)/kT)$, (4)

where $W_0$ and μ are constants, $k$ is the Boltzmann constant, and $T$ is temperature,

$X = K + L\log(1+Mt)$, (5)

where $K$, $L$, and $M$ are constants. [1]

The Wagner theory [18, 19] might be applicable to thick films ($\gtrsim$1 μm), which postulates that the film thickness $X$ grows proportional to $t^{1/2}$ (parabolic kinetics), where $t$ is time, when the film growth rate is limited by diffusion across the film. This happens when the growth rate $dX/dt$ is proportional to the reciprocal of film thickness ($1/X$). When the concentration difference of a diffusing species is always the same between two sides of the film, the diffusion flux is proportional to $1/X$, and the film growth rate is also proportional to $1/X$ assuming Fick's law.

Computational studies that directly observe spontaneous oxide film formation are difficult. Supercells containing both the metal slab and $O_2$ gas should contain at least on the order of $10^3$ atoms. Molecular dynamics (MD) simulations using density functional theory (DFT) are not practical for supercells of this size.

The ReaxFF reactive force-field interatomic potential (ReaxFF) [20] is a highly effective method for specific systems [21], and examples of applications are nano-$TiO_2$ surface hydrogenation [22], oxidation of Ni groove defects [23] and Fe-Ni alloy nanowires [24].

A very large number of atoms can be handled with ReaxFF. An Al nanoparticle oxidation study involved 7 nm (~$10^4$ atom) nanoparticles [25]. A computational study by Zhang [26] showed that spherical Al nanoparticles exposed to $O_2$ gas can oxidize in three morphologies, which are a core-shell cluster where oxides form on the surface of the nanoparticle, a core-chain structure where chains of $Al_4O$ raises from the nanoparticle surface and are subsequently oxidized, and splitting into small independent oxides. Diverse oxidation behavior is possible in nanoparticles because there are highly reactive atoms at the surface, for instance, at corner and edge sites. A ReaxFF study of Mg nanoparticle oxidation showed that a spherical Mg nanoparticle can ultimately oxidize to rocksalt structure MgO.[27]

In this study, brute-force MD results of Al, Mg, and Ti surface oxidation were conducted with a supercell containing a few thousand atoms. The PFP [28] universal potential was in this study without fine tuning, demonstrating successful descriptions of the same potential on different representative systems without any prior potential tuning.

## 2. Methods

### 2.1 General MD settings

MD simulations were conducted using the commercially available Matlantis package from Preferred Networks with their universal PFP [28] version 6.0.0, a NNP trained on VASP code [29, 30] results based on Perdew-Burke-Ernzerhof (PBE) generalized gradient approximation (GGA) to DFT [31]. Diverse compounds consisting of any of 72 elements may be modeled using the same potential trained on 42 million training data. The PFP is available as a fixed pretrained potential that cannot be modified by the user. Fine-tuning potentials to specific systems is outside the scope of this work.

Temperatures considered were 300, 500, or 700 K; the melting temperatures of Al, Mg, and Ti are 933, 923, and 1941 K, respectively, at standard atmospheric pressure (1 atm).

### 2.2 Metal model generation

The VASP PBE lattice parameters are $a$=4.0401 Å for fcc Al, $a$=3.1925 Å and $c$=5.1782 Å for hcp Mg, and $a$=2.9225 Å and $c$=4.6291 Å for hcp Ti, respectively. The fcc (111) and hcp (0001) surfaces are the most close-packed, and therefore have the lowest surface energies. The distance between adjacent Al (111), Mg (0001), and Ti (0001) layers are $a/\sqrt{3}$=2.3326 Å, $c/2$=2.5891 Å, and $c/2$= 2.3145 Å, respectively. Orthogonal supercells containing 56 atoms in each layer, as in Ref. [32], and 108 Al, 96 Mg, or 108 Ti layers were constructed, with atoms at fractional $z$-coordinates of $(0.5+k)/108$, $(0.5+k)/96$, or $(0.5+k)/108$, respectively, where $k$ is an integer. The supercell size for MD simulations

was $a$=19.793 Å, $b$=20.000 Å, and $c$=251.920 Å for Al, $a$=22.118 Å, $b$=22.347 Å, and $c$=248.556 Å for Mg, and $a$=20.248 Å, $b$=20.457 Å, and $c$= 249.969 Å for Ti, respectively. Atoms with $z$-coordinates between $1/8<z<7/8$ were removed, resulting in 1568 Al atoms, 1344 Mg atoms, or 1568 Ti atoms. There were initially no atoms at $z$=0. A short MD simulation of 5000 steps at 2 fs intervals (10 ps) was conducted to attain equilibrium of the metal slab at the desired temperature.

Convergence of temperature, energy, and outermost $z$-coordinate results are discussed in section S1 in the supplementary material using Figs. S1-4. We stress that the Nosé–Hoover [33, 34] thermostat is very inappropriate when simulating metal slabs because the metal slab thickness oscillates even after 200 ps, while the thickness equilibrates within 10 ps with Langevin dynamics.

**2.3 Initial $O_2$ gas distribution generation**

The initial $O_2$ gas distribution fills the space between the metal slabs and also serves as an $O_2$ reservoir that replenishes $O_2$ consumed to form the oxide film. A reasonable $O_2$ gas distribution was obtained by heating $O_2$ molecules initially placed in an ordered pattern. The gas should be trapped between two walls, and the walls were modeled using inert Ar atoms that were packed close to each other. Motion of Ar atoms were suppressed by using an extremely large atom mass of $10^9$ atomic mass units. The role of the Ar wall is to make an $O_2$ gas confined in one direction. This artificial wall is removed and replaced with the metal slab in oxidation calculations.

The O atoms were initially positioned at fractional coordinates ($i$/4, $j$/4, ($k$+0.5)/24±0.0024) with integers $0\leq i<4$, $0\leq j<4$, and $3\leq k<21$. The initial O-O bond distance was roughly 1.2 Å. The Ar walls were placed at $z$=0.132 and 1-0.132 for Al and $z$=0.128 and 1-0.128 for Mg and Ti, which are the outermost metal positions after MD simulations of the metal slabs at 700 K. An example of atom positions before and after such a MD simulation is shown in Fig. S5.

The initial $O_2$ gas pressure of the gas, expressed in units of atm, was crudely estimated at 160, 260, and 370 for Al, 130, 210, and 300 for Mg, and 150, 250, and 350 for Ti at 300, 500, and 700 K, respectively, assuming that the $O_2$ is an ideal gas and that the gas exists in a space of 3/4 of the supercell. The pressures are different because the same number of $O_2$ molecules was inserted in the supercells, regardless of their lattice parameters, and the pressure is proportional to the temperature in this approximation.

### 2.4 Oxidation simulation

MD calculations were carried out with a 2 fs time step for 2 ns ($10^6$ steps), and atom positions were recorded every 500 steps (1 ps). The initial model with both metal and $O_2$ was obtained by simply superimposing the atom positions of the metal slab and $O_2$ reservoir.

Typical fixed number of atoms, volume, and temperature (NVT) simulations cannot appropriately simulate surface oxidation as in this study. The $O_2$ gas concentration between the slabs, which is the $O_2$ gas pressure, should decrease as an NVT simulation progresses because more and more $O_2$ molecules will be absorbed in the metal while O is not released from the metal oxide. This results in a very low $O_2$ pressure region near the surface, at least in very early stages of oxidation. Adding more $O_2$ molecules near the surface can accelerate the oxidation process.

Addition of molecules was attempted every 100 steps by copying $O_2$ molecules from the reservoir to the supercell at certain positions if there were no $O_2$ molecules around. The supercell was separated into a 3×3×40 grid (6-7 Å each side). When there were no molecules in a grid, a molecule was picked from the reservoir and added in the center of the grid. The orientation and velocity of the molecule was not changed from the reservoir, and only the center of mass was changed. This procedure results in a continuous increase of the $O_2$ pressure between the slabs because molecules can be inserted far away from the surface and not just near the surface.

A ReaxFF MD study exposed an Al slab to a $O_2$ gas flow with an initial velocity toward one surface. $O_2$ molecules were added at a rate of $1.4\times10^{-2}$ molecules·$ps^{-1}$·$Å^{-2}$. $O_2$ molecules that did not react with the surface were removed before new O were introduced to the supercell. This procedure made possible observation of how O and Al react at low O coverage on a 40 Å × 40 Å surface. [35]. We are interested in formation of a native film within the constraints of a reasonable the simulation time, thus we aimed to maintain a high O pressure to provide more chances for surface oxidation.

Adsorption and splitting of $O_2$ at the metal surface should be a very exothermic reaction, at least at the start of the simulation, locally releasing much heat at the surface. In reality, this heat should transfer deep into the macroscopic metal particle, but this is not possible in the relatively small supercell of the MD simulation. Moreover, heat transfer by electrons cannot be modeled using classical MD simulations, including NNP-MD. This study separated the molecule into three regions and used different dynamics in each region. Langevin dynamics were used within the metal region, which was defined as between *z*-coordinates of the outermost metal atoms. The fixed number of atoms, volume, and energy (NVE) velocity Verlet algorithm was used in the interface region extending 15 Å into the $O_2$ gas from the metal region boundaries. The canonical sampling through velocity rescaling (CSVR) [36] was used elsewhere, which is the gas region. The friction term in Langevin dynamics was 0.03 $fs^{-1}$ and the CSVR τ constant was 1000 fs. Convergence of the metal and O temperatures between 0 and 20 ps is shown in Fig. S6. The O temperature overshoots at the start of the simulation but converges after 5-15 ps, while there is no overshoot for the metal. The amount of O in the MD supercell increases with time, as is discussed in section S2 in the supplementary material using Fig. S7.

Snapshots of the supercell containing Mg and O after 1 ps and 2000 ps at 300 K are given in Figs. 1(a,b), respectively. The metal, interface, and gas regions, together with their thermostats, as shown in Fig. 1(b). Note that the metal-region boundary is located at the outermost metal atom. A thick oxide layer has formed by 2000 ps, and the $O_2$ gas density visibly increases with simulation time.

### 2.5 Pair distribution function (PDF)

The PDF is a measure of the short-range order of atoms. Computationally, the PDF, $g(r)$, can be expressed as a superimposition of delta functions, but such a PDF is difficult to visualize and compare with experiments. The delta functions were broadened with a Gaussian with parameter σ to define the non-normalized PDF in this study as a function of distance $r$ as

$$g(r) = \sum_{ij} \frac{1}{4\pi r_{ij}^2} \left( \frac{1}{2\pi\sigma} \right) \exp\left( -\frac{(r - r_{ij})^2}{2\sigma^2} \right), \quad (6)$$

where $r_{ij}$ is the distance between atoms $i$ and $j$. The summation is limited to a certain pair of species in this study, namely the metal and O. The atom $i$ is within a supercell, while the other atom $j$ may be outside the supercell. PDFs were calculated within an arbitrary cutoff distance. The broadening reflects the thermal fluctuation near an equilibrium position at a finite temperature.

A normalized PDF, $g'(r)$, is defined as

$$g'(r) = c\sum_{ij}\frac{1}{r_{ij}^2}\exp\left(-\frac{\left(r-r_{ij}\right)^2}{2\sigma^2}\right). \quad (7)$$

This quantity is plotted in this study. Here, $c$ is an arbitrary scaling constant such that the maximum value of $g'(r)$ is 1. The shape of the PDF is important, especially the peak positions, but usually the absolute value is not. The σ in this study was 0.1 Å.

## 3. Results and discussion

### 3.1 Oxide film thickening with time

Fig. 2 shows snapshots of the slab near the $z = 0$ end for Al, Mg, and Ti at 1 ps, 20 ps, 100 ps, 500 ps, and 2000 ps for 300 K and 100 ps, 500 ps, and 2000 ps for 500 K and 700 K, respectively. The full supercells of snapshots in Fig. 2 are shown in Fig. S8. In the initial stage up to 20 ps, O molecules that approach the metal surface are adsorbed on the clean metal, forming vacuum at the surface (Fig. 2(b,m,x)). The vacuum disappears at 100 ps because O is sufficiently supplied from the gas region and the surface oxide suppresses adsorption of more O atoms. Al and Ti oxide films gradually thicken at the surface (Figs. 2 (a-k) and Figs. 2 (w-gg), respectively), while the Mg oxide film quickly thickens into the metal (Fig. 2 (l-v)).

Reax-FF simulations of an Al slab and $O_2$ gas with a density of 0.12 or 0.24 g/cm$^3$ at 300 K stated that the slab thickness from the former is about half of the latter after 1 ns. However, most of the $O_2$ gas had oxidized after 1 ns.[37] This shows the importance of providing sufficient $O_2$ gas during the entire simulation to obtain the maximum possible slab thickness.

An environmental transmission electron microscopy study [38] showed how the oxide film grows on the Al (100) surface under an $O_2$ gas pressure of $3\times10^{-5}$ Torr. Islands of a ~1 nm thick amorphous oxide had formed near kinks by 50 s after O was introduced. The first oxide monolayer formed between the established oxide at 54 s. After subsequent layer-by-layer oxidization, the bridging oxide between the islands became ~1 nm thick, which is almost the same as the islands. The bridging oxide was roughly two times thicker toward vacuum compared to toward the Al bulk when measured from the initial Al surface. The calculated Al oxide thickness for 300 K is thicker toward the $O_2$ gas region compared to the Al bulk, consistent with the experimental results.

The native oxide on a Mg-2Al-0.1Ca alloy, which is a Mg alloy with 2 and 0.1 wt% of Al and Ca, respectively, is a 3-4 nm thick layer of MgO (111). A high Al content region forms at the metal-oxide interface, hence the native oxide formation behavior could be different between Mg and Al-containing Mg alloys. [39]

As a more quantitative analysis, the supercell was separated into $N$ regions along the $z$-direction, where $N$ is 108 for Al and Ti and 96 for Mg. A metal layer is at the middle of each region, in the $z$-direction, in the cleaved metal slab prior to any relaxation. The temporal change in the number of metal and O atoms in each region indicates how the oxide layer thickness changes, and the metal to O ratio gives the approximate stoichiometry of the oxide layer.

Fig. 3(a-c) provides histograms of atoms in each region for select time steps over all temperatures. The metal to O ratio is approximately 1:1, 2:3, and 1:2 for Mg, Al, and Ti, respectively, suggesting that the stoichiometry of the passivation layer is approximately MgO, $Al_2O_3$, and $TiO_2$. However, there is clearly more O than Mg at the outermost surface for 100 ps and onwards in 300 K only. The $z$-coordinates in Fig. 3 are in fractional coordinates. Fig. 3(d-f) are histograms, at 500 ps, of the metal-O coordination number (CN) of atoms. Atoms with zero CN are not shown, and a non-zero CN indicates atoms forming an oxide in the interface region. Metal and O atoms tend have small CN at small and large $z$-coordinates, in the interface region, respectively. Interestingly, a large fraction of Mg at large $z$, which is very near the gas region, has a higher CN than within the interface region. The closed-shell $Mg^{2+}$ prefers six-fold coordination over four-fold. The abundance of open space and O atoms allows formation of an oxide at the outermost surface with high coordination Mg and low coordination O, resulting in a Mg:O stochiometry that is much O-rich than 1:1.

Figs. 4 (a,c,e) shows how the $z$-coordinates ($0<z<1/2$) of the outermost metal (largest) and O (smallest) change with simulation time. The difference between the two z-coordinates is not necessarily the best measure of the oxide layer because the coordinates are obtained from a single atom, not some weighted average of atoms at the reaction front.

The slab thickness, defined as the difference between outermost metal and O positions, is plotted against the logarithm of temperature in Figs. 4(b,d,f). A clear direct logarithmic growth law is found for all metals and temperatures (the 700 K Mg result is valid for up to ~0.8 ps only because there is no pure Mg afterwards), showing that the oxide formation is governed by an activation energy that linearly increases with film thickness, consistent with Cabrera-Mott theory. The slopes in Figs. 4(b,d,f) are, in Å/decade, 3.7, 4.0, and 6.6 at 300, 500, 700 K, respectively, for Al (Fig. 4(b)), 9.4 and 10.2 at 300 and 500 K, respectively, for Mg (Fig. 4(d)), and 5.7, 7.6, and 3.7 at 300, 500, 700 K, respectively, for Ti (Fig. 4(f)). As a very crude estimate, a slope of 5 Å/decade over 30 years (~$10^{18}$ ns) under an extremely high oxygen pressure of ~$10^2$ atm order results in a growth of 9 nm. Assuming that that the slab growth rate is proportional to the $O_2$ partial pressure because the growth rate is limited by the number of gas molecules approaching the surface within a certain time, the film thickness is decreased by roughly the growth rate times three decades because the $O_2$ pressure in the simulation is ~$10^3$ times higher than at ambient conditions. Native films exceeding few ten nm are not expected for the considered systems unless there is an additional mechanism that accelerates the surface oxide growth.

### 3.2 PDF analysis

Fig. 5 shows the metal-O PDFs after the simulations (2 ns, after 0.5 ns for Mg-O at 700 K only). The PDFs are almost the same over all temperatures (Figs. 5(a,c,e)). In Al and Mg oxides, short-range ordering involving the nearest neighbor (NN) only is found since all peaks other than a single peak at ~2 Å are very broad. In contrast, Ti-O peaks are additionally found at ~3.7 Å and ~4.6 Å, indicating order over a longer range than Al-O and Mg-O.

The peak positions are compared to PDFs of bulk oxides at 0 K. Al atoms in the corundum structure α-$Al_2O_3$ have six-fold coordination, while Al in γ-$LiAlO_2$ (space group $P4_12_12$, number 92) [40] are four-fold coordinated. Mg atoms in rocksalt and zincblende structure MgO are six- and four-fold coordinated, respectively. Ti in the rutile and anatase $TiO_2$ structures are six-fold coordinated, while Ti atoms in β-$Ba_2TiO_4$ (space group reported as $P2_1/n$, standard setting $P2_1/c$, number 14)[41] are four-fold coordinated.

The NN peak position for Al-O, 1.82 Å, is much closer to γ-$LiAlO_2$ (1.78 Å) than α-$Al_2O_3$ (1.92 Å). The Al of the film is considered ~4.4-fold coordinated if the average coordination linearly changes with NN peak position. The Mg of the film is ~4.3-fold coordinated based on the Mg-O NN peaks of the film, zincblende MgO, and rocksalt MgO, which are 2.02, 2.00, and 2.13 Å, respectively, while Ti of the film is ~5.2-fold coordinated based on the Ti-O NN peaks of the film, β-$Ba_2TiO_4$, rutile $TiO_2$, and anatase $TiO_2$ (1.92, 1.84, 1.98, and 1.74 Å, respectively).

Experimental evidence shows that $Ti_2O_3$ may form at the surface when the $O_2$ pressure is low and the temperature is high because the O solubility is fast enough, however, $Ti_2O_3$ would not form in the simulations in the present simulations because O is abundant.[42]

## 4. Conclusions

Growth of oxide films on Al, Mg, and Ti slabs was modeled through MD simulations using a universal NNP. Langevin and CSVR thermostats were used for the metal and gas regions, respectively, and NVE dynamics for the interface region. $O_2$ gas was supplied to the gas region to compensate for O incorporated in the metal to form the oxide. Al and Ti formed a 1-2 nm thick film within a few ps of the simulation with $O_2$ gas pressure of $\sim 10^6$ atm, while the Mg oxide film thickness continued to increase proportionally to the logarithm of simulation time. Al and Ti oxides grow at the surface, while Mg oxide grows into the Mg metal. Al, Mg, and Ti are approximately 4.4, 4.3, and 5.2-fold coordinated in the oxide, respectively. In contrast to Langevin dynamics, NH dynamics is unsuitable for NVT MD simulations of at least Al, Mg, and Ti slabs because oscillations of the slab thickness do not subside.

## Supplementary Material

Four figures comparing Langevin and NH dynamics on metal slabs and three figures of MD snapshots containing metal and O.

## Acknowledgments

The VESTA code [43] was used to draw Figures 1, 2, S5, and S8.

## Conflict of Interest statement

Taisuke Hasegawa is an employee of Matlantis Corporation, the provider of the Matlantis platform used in this study.

## Data availability statement

The data that supports the findings of this article cannot be made publicly available because the data is too large and no suitable public repository exists for hosting in this field of study. The software used for this research is owned by a third party and cannot be publicly shared.

## Author Contributions

Y.H: Conceptualization, Data Curation, Formal Analysis, Investigation, Methodology, Project Administration, Resources, Software, Supervision, Validation, Visualization, Writing/Original Draft Preparation, Writing/Review & Editing

T.H: Data Curation, Formal Analysis, Investigation, Methodology, Software, Writing/Original Draft Preparation, Writing/Review & Editing

**References**

[1] P. Schmuki, From Bacon to barriers: a review on the passivity of metals and alloys, J. Solid State Electrochem., 6 (2002) 145-164.

[2] S. Li, L. Yang, J. Christudasjustus, N.R. Overman, B.D. Wirth, M.L. Sushko, P. Simonnin, D.K. Schreiber, F. Gao, C. Wang, Selective atomic sieving across metal/oxide interface for super-oxidation resistance, Nat. Commun., 15 (2024) 6149.

[3] B.E. Ibrahimi, J.V. Nardeli, L. Guo, An Overview of Corrosion, in: Sustainable Corrosion Inhibitors I: Fundamentals, Methodologies, and Industrial Applications, American Chemical Society, 2021, pp. 1-19.

[4] Z. Zhou, X. Chen, D. Wu, D. Zhu, J. Chen, X. Sun, M. Li, C. Cai, J.C. Yang, G. Zhou, In situ electron microscopy: atomic-scale dynamics of metal oxidation and corrosion, npj Materials Degradation, 9 (2025) 28.

[5] J. Gorobez, B. Maack, N. Nilius, Growth of Self-Passivating Oxide Layers on Aluminum—Pressure and Temperature Dependence, physica status solidi (b), 258 (2021) 2000559.

[6] N. Cai, G. Zhou, K. Müller, D.E. Starr, Tuning the Limiting Thickness of a Thin Oxide Layer on Al(111) with Oxygen Gas Pressure, Phys. Rev. Lett., 107 (2011) 035502.

[7] L.P.H. Jeurgens, W.G. Sloof, F.D. Tichelaar, C.G. Borsboom, E.J. Mittemeijer, Determination of thickness and composition of aluminium-oxide overlayers on aluminium substrates, Appl. Surf. Sci., 144-145 (1999) 11-15.

[8] L.P.H. Jeurgens, W.G. Sloof, F.D. Tichelaar, E.J. Mittemeijer, Growth kinetics and mechanisms of aluminum-oxide films formed by thermal oxidation of aluminum, J. Appl. Phys., 92 (2002) 1649-1656.

[9] R. Nakamura, D. Tokozakura, H. Nakajima, J.-G. Lee, H. Mori, Hollow oxide formation by oxidation of Al and Cu nanoparticles, J. Appl. Phys., 101 (2007) 074303.

[10] D. Starodub, T. Gustafsson, E. Garfunkel, The reaction of O2 with Al(110): a medium energy ion scattering study of nano-scale oxidation, Surf. Sci., 552 (2004) 199-214.

[11] S.V. Metelev, N.K. Pleshanov, A. Menelle, V.M. Pusenkov, A.F. Schebetov, Z.N.

Soroko, V.A. Ul'yanov, The study of oxidation of thin metal films by neutron reflectometry, Physica B: Condensed Matter, 297 (2001) 122-125.

[12] J.M. Jáquez-Muñoz, C. Gaona-Tiburcio, C.T. Mendez-Ramirez, M.G. Carrera-Ramirez, M.A. Baltazar-Zamora, G. Santiago-Hurtado, M. Lara-Banda, F. Estupiñan-Lopez, D. Nieves-Mendoza, F. Almeraya-Calderon, Corrosion of Anodized Titanium Alloys, Coatings, 14 (2024) 809.

[13] S.F. Jr., J.C. Galván, A. Pardo, M.C. Merino, R. Arrabal, Native Air-Formed Oxide Film and its Effect on Magnesium Alloys
Corrosion, The Open Corrosion Journal, 2010 (2010) 80-91.

[14] A. Fattah-alhosseini, R. Chaharmahali, A. Askari, S. Alizad, M. Kaseem, Unraveling the impact of purification and alloying elements on corrosion performance and passivation of magnesium alloys, Journal of Magnesium and Alloys, 12 (2024) 4808-4827.

[15] T. Wu, K. Zhang, Corrosion and Protection of Magnesium Alloys: Recent Advances and Future Perspectives, Coatings, 13 (2023) 1533.

[16] N. Cabrera, N.F. Mott, Theory of the oxidation of metals, Rep. Prog. Phys., 12 (1949) 163.

[17] J.D. Baran, H. Grönbeck, A. Hellman, Mechanism for Limiting Thickness of Thin Oxide Films on Aluminum, Phys. Rev. Lett., 112 (2014) 146103.

[18] A. Atkinson, Transport processes during the growth of oxide films at elevated temperature, Rev. Mod. Phys., 57 (1985) 437-470.

[19] C. Wagner, Beitrag zur Theorie des Anlaufvorgangs, Z. Phys. Chem., 21B (1933) 25-41.

[20] T.P. Senftle, S. Hong, M.M. Islam, S.B. Kylasa, Y. Zheng, Y.K. Shin, C. Junkermeier, R. Engel-Herbert, M.J. Janik, H.M. Aktulga, T. Verstraelen, A. Grama, A.C.T. van Duin, The ReaxFF reactive force-field: development, applications and future directions, NPJ Comput. Mater., 2 (2016) 15011.

[21] X. Zhang, P. Zheng, Y. Ma, Y. Jiang, H. Li, Atomic-scale understanding of oxidation mechanisms of materials by computational approaches: A review, Materials & Design, 217 (2022) 110605.

[22] P. Zheng, L. Zhang, X. Zhang, Y. Ma, J. Qian, Y. Jiang, H. Li, Hydrogenation of TiO2 nanosheets and nanoparticles: typical reduction stages and orientation-related anisotropic disorder, Journal of Materials Chemistry A, 9 (2021) 22603-22614.

[23] Y. Ma, D. Zhang, P. Zheng, J. Qian, Y. Wang, Y. Jiang, H. Li, Oxidation of nickel with groove defects: Cation vacancies expansion mechanism and directional layer-by-layer oxidation, Appl. Surf. Sci., 593 (2022) 153384.

[24] M. Fu, R. Bai, X. Zhang, Q. Liu, Z. Li, C. Che, S. Cheng, L. Zhang, W. Wu, H. Li, Oxidation mechanism transition and phase separation in Fe-Ni alloy Nanowires: Bridging experiments with molecular dynamics, Appl. Surf. Sci., 715 (2026) 164556.

[25] Y. Zhang, S. Wang, Z. Zhou, H. Zhao, C. Tang, Y. Gao, Study on the effect of particle size on early oxidation characteristics of aluminum nanoparticles, Powder Technol., 400 (2022) 117227.

[26] X. Zhang, C. Fu, Y. Xia, Y. Duan, Y. Li, Z. Wang, Y. Jiang, H. Li, Atomistic Origin of the Complex Morphological Evolution of Aluminum Nanoparticles during Oxidation: A Chain-like Oxide Nucleation and Growth Mechanism, ACS Nano, 13 (2019) 3005-3014.

[27] F. Fiesinger, D. Gaissmaier, M. van den Borg, J. Beßner, A.C.T. van Duin, T. Jacob, Development of a Mg/O ReaxFF Potential to describe the Passivation Processes in Magnesium-Ion Batteries, ChemSusChem, 16 (2023) e202201821.

[28] S. Takamoto, C. Shinagawa, D. Motoki, K. Nakago, W. Li, I. Kurata, T. Watanabe, Y. Yayama, H. Iriguchi, Y. Asano, T. Onodera, T. Ishii, T. Kudo, H. Ono, R. Sawada, R. Ishitani, M. Ong, T. Yamaguchi, T. Kataoka, A. Hayashi, N. Charoenphakdee, T. Ibuka, Towards universal neural network potential for material discovery applicable to arbitrary combination of 45 elements, Nat. Commun., 13 (2022) 2991.

[29] G. Kresse, J. Furthmüller, Efficient iterative schemes for ab initio total-energy calculations using a plane-wave basis set, Phys. Rev. B, 54 (1996) 11169-11186.

[30] G. Kresse, D. Joubert, From ultrasoft pseudopotentials to the projector augmented-wave method, Phys. Rev. B, 59 (1999) 1758-1775.

[31] J.P. Perdew, K. Burke, M. Ernzerhof, Generalized Gradient Approximation Made Simple, Phys. Rev. Lett., 77 (1996) 3865-3868.

[32] Y. Hinuma, Generation of almost rectangular, square, and hexagonal two-dimensional supercells, Science and Technology of Advanced Materials: Methods, 4 (2024) 2300254.

[33] S. Nosé, A unified formulation of the constant temperature molecular dynamics methods, J. Chem. Phys., 81 (1984) 511-519.

[34] W.G. Hoover, Canonical dynamics: Equilibrium phase-space distributions, Phys. Rev. A, 31 (1985) 1695-1697.

[35] Y. Kim, J. Choi, Oxide growth characteristics on Al (100), (110), and (111) surfaces: A chemo-mechanical evaluation, Materials Today Communications, 26 (2021) 102006.

[36] G. Bussi, D. Donadio, M. Parrinello, Canonical sampling through velocity rescaling, J. Chem. Phys., 126 (2007) 014101.

[37] M.E. Trybula, P.A. Korzhavyi, Atomistic Simulations of Al(100) and Al(111) Surface

Oxidation: Chemical and Topological Aspects of the Oxide Structure, J. Phys. Chem. C, 123 (2019) 334-346.

[38] L. Nguyen, T. Hashimoto, D.N. Zakharov, E.A. Stach, A.P. Rooney, B. Berkels, G.E. Thompson, S.J. Haigh, T.L. Burnett, Atomic-Scale Insights into the Oxidation of Aluminum, ACS Applied Materials & Interfaces, 10 (2018) 2230-2235.

[39] D. Neuß, I.E. McCarroll, S. Zhang, E. Woods, W.J. Delis, L. Tanure, H. Springer, S. Sandlöbes, J. Yang, M. Todorova, D. Zander, C. Scheu, J.M. Schneider, M. Hans, High-resolution chemical and structural characterization of the native oxide scale on a Mg-based alloy, Corros. Sci., 227 (2024) 111776.

[40] M. Marezio, The crystal structure and anomalous dispersion of [gamma]-LiAlO2, Acta Crystallogr., 19 (1965) 396-400.

[41] K.K. Wu, I.D. Brown, The crystal structure of [beta]-barium orthotitanate, [beta]Ba2TiO4, and the bond strength-bond length curve of TiO, Acta Cryst. B, 29 (1973) 2009-2012.

[42] I. Vaquila, L.I. Vergara, M.C.G. Passeggi, R.A. Vidal, J. Ferrón, Chemical reactions at surfaces: titanium oxidation, Surf. Coat. Technol., 122 (1999) 67-71.

[43] K. Momma, F. Izumi, VESTA 3 for three-dimensional visualization of crystal, volumetric and morphology data, J. Appl. Crystallogr., 44 (2011) 1272-1276.

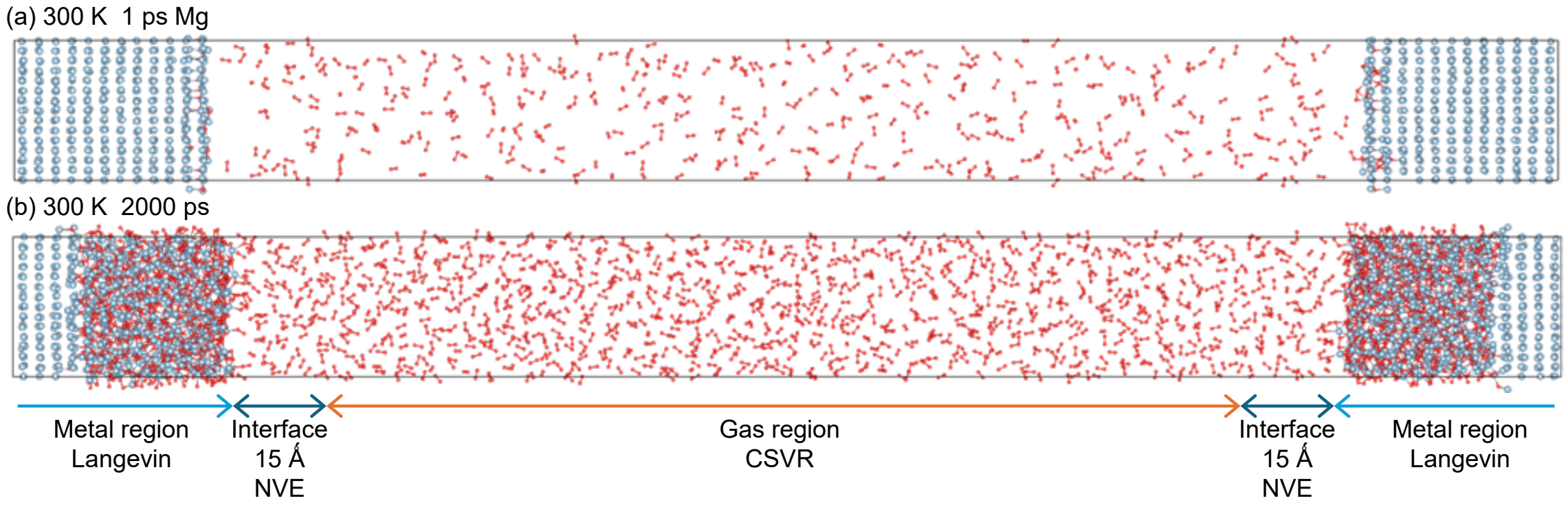

Fig. 1. Atom positions of snapshots when the slab is Mg and the temperature is 300 K. The time step is (a) 1 ps and (b) 2000 ps, respectively. The metal, interface, and gas regions, together with their thermostats, are shown in (b).

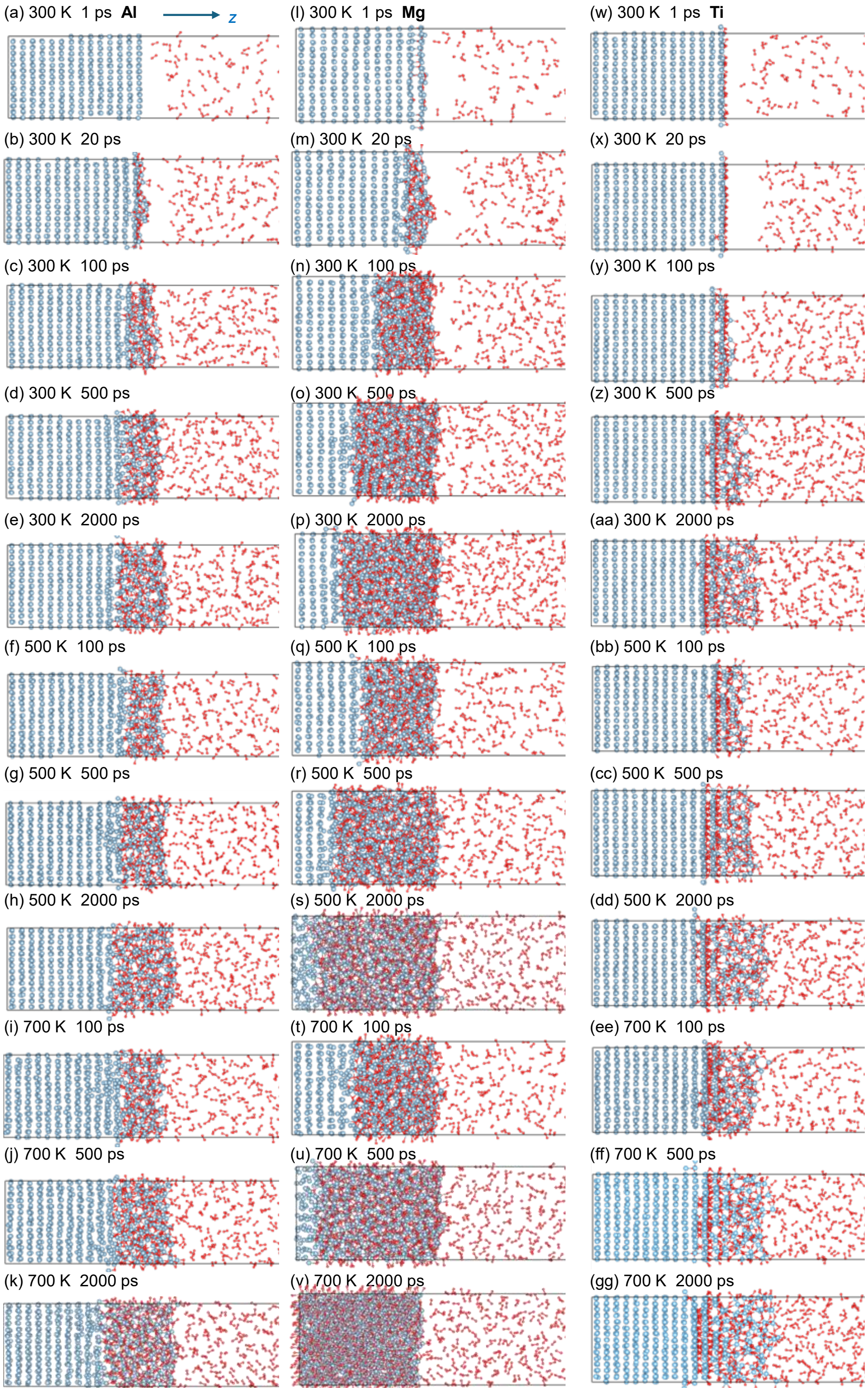

Fig. 2. Atom positions of snapshots, for *z*-coordinates between 0 and roughly 0.25, at selected temperatures and simulation time when the slab is (a-k) Al, (l-v) Mg, or (w-gg) Ti, respectively. Large blue circles indicate (a-k) Al, (l-v) Mg, or (w-gg) Ti. Small red circles represent O.

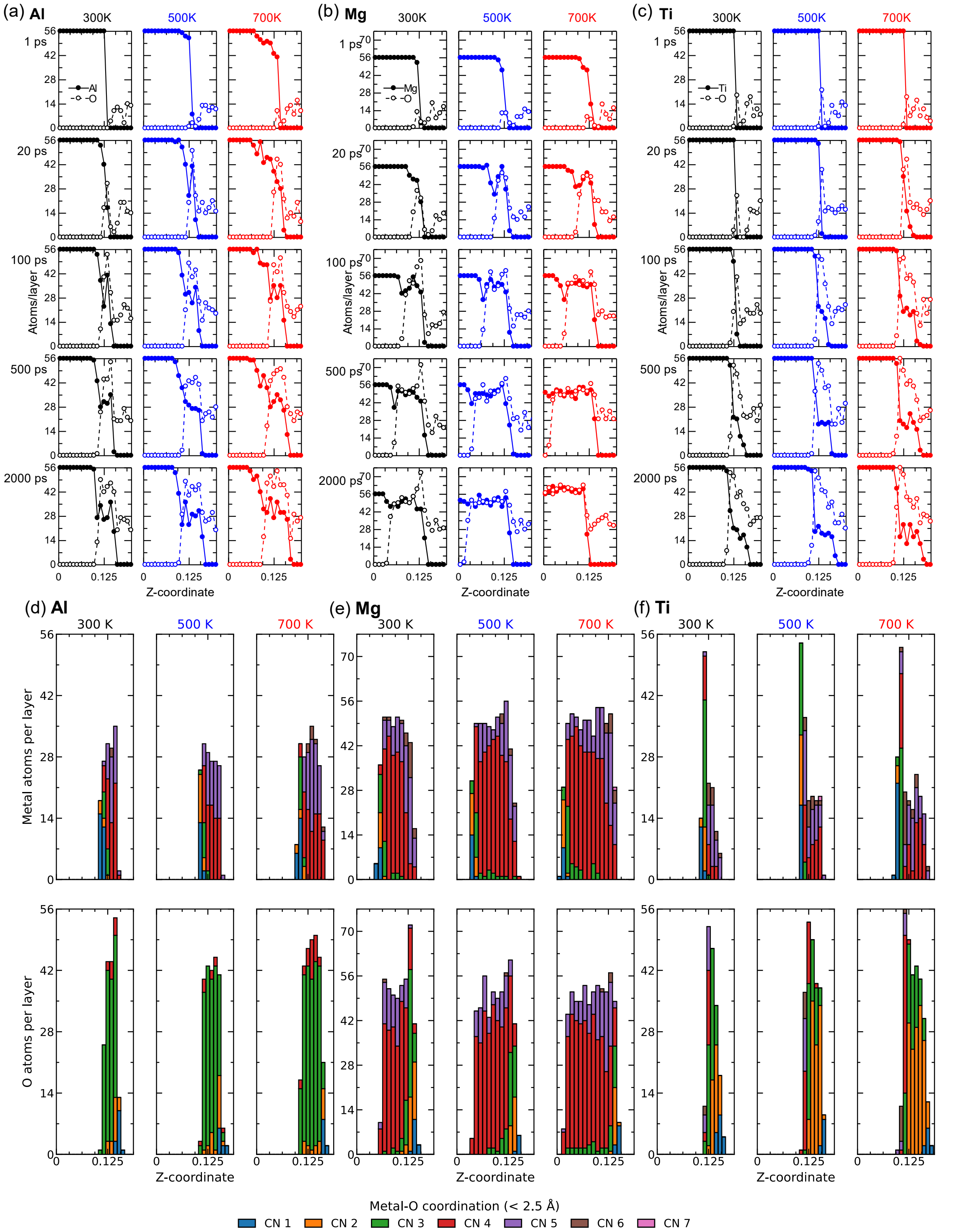

Fig. 3. (a-c) Histograms of the number of O and (a) Al, (b) Mg, or (c) Ti atoms in the supercell with (a,c) 108 or (b) 96 bins (layers) along the z-axis. (d-f) Histograms of (d) Al, (e) Mg, or (f) Ti and O atoms in each layer that have a non-zero coordination number (CN). The *z*-coordinates in this figure are in fractional coordinates.

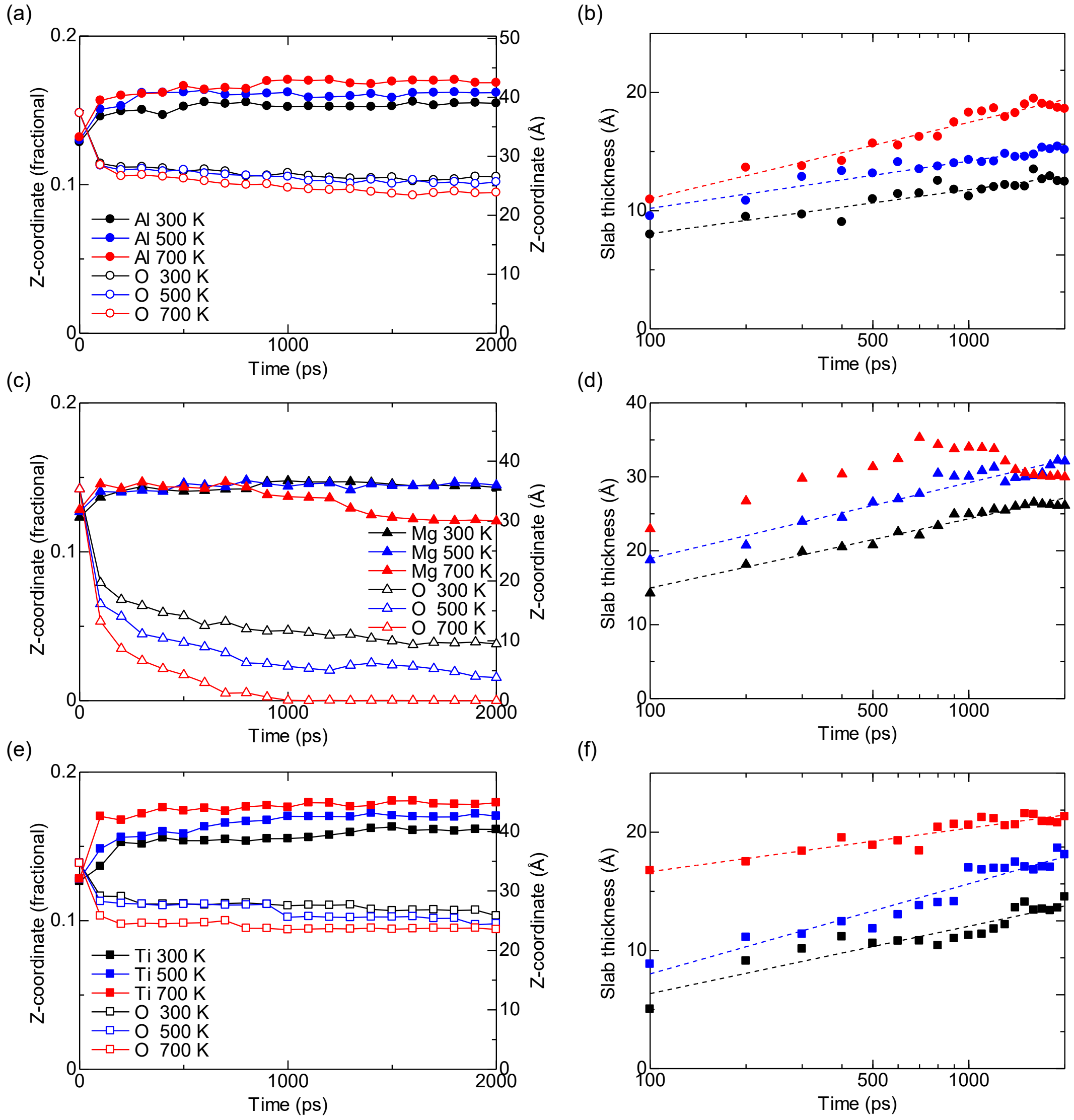

Fig. 4. (a,c,e) Largest *z*-coordinate of (a) Al, (c) Mg, and (f) Ti atoms, respectively, and smallest *z*-coordinate of O atoms in the range $0<z<1/2$ at select simulation time steps. (b,d,f) Slab thickness defined as the difference between the largest *z*-coordinate of (b) Al, (d) Mg, and (f) Ti atoms, respectively.

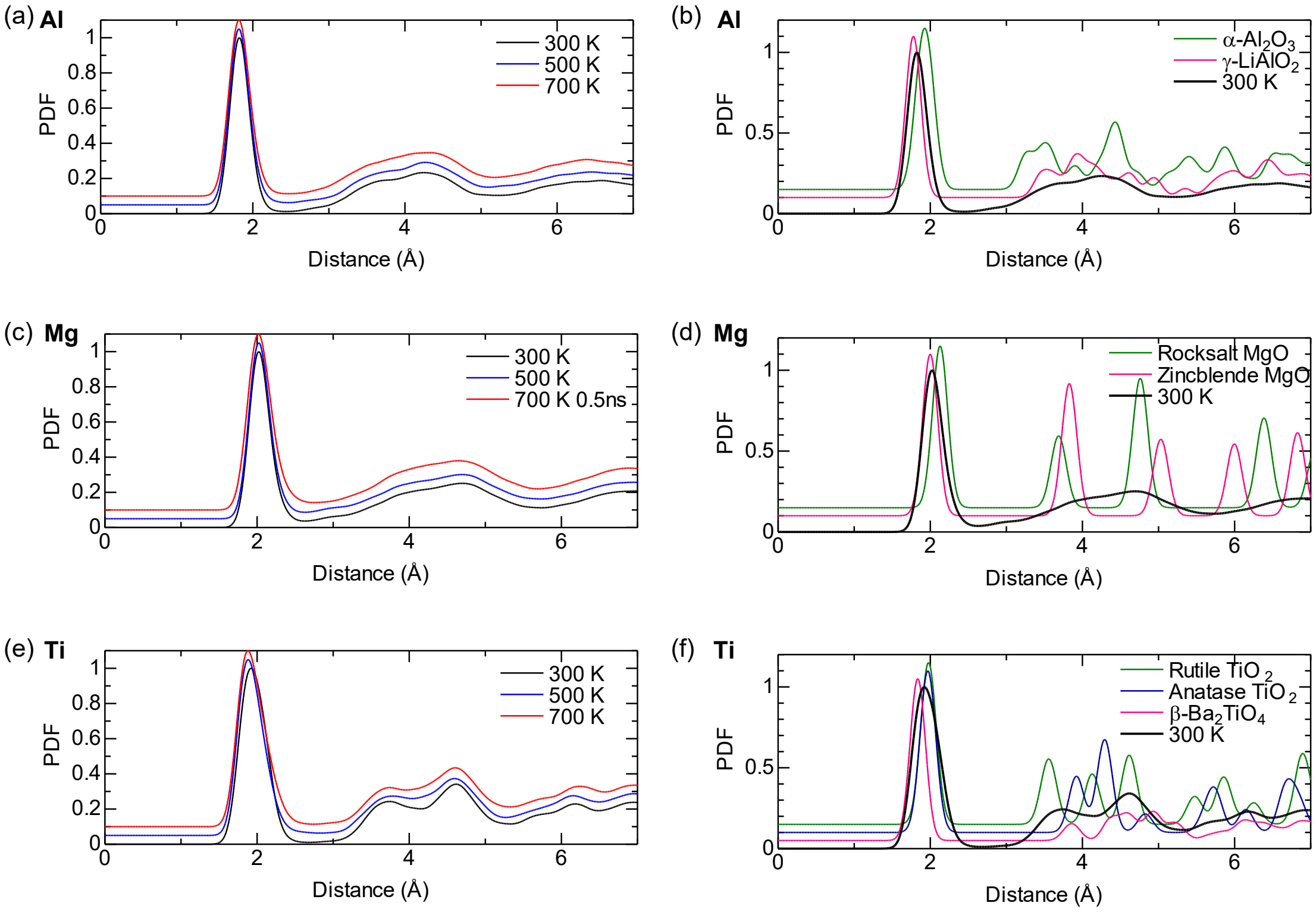

Fig. 5. (a,b) Al-O, (c,d) Mg-O, and (e,f) Ti-O PDFs after MD simulations over 2 ns. 300 K, 500 K, and 700 K results are compared in (a,c,e), and the 300 K results are compared to reference oxides at 0 K in (b,d,f). The Mg-O PDFs after 0.5 ns are shown instead of 2 ns in (c) because there is no oxide-Mg metal interface after 2 ns. PDFs are shifted vertically in (a,c,e) to avoid overlapping of curves.